# Superradiation, Superdirectivity and Efficiency Boosting in Coherently Driven Antennas

Jian Wen Choong[1] and Alex Krasnok[1*]

[1]*Photonics Initiative, Advanced Science Research Center, City University of New York, NY 10031, USA*

*To whom correspondence should be addressed: akrasnok@gc.cuny.edu*

## Abstract

Antennas, i.e., elements transforming localized or waveguide modes into freely propagating fields and vice versa, are vital components for wireless technologies across the entire spectrum of electromagnetic waves, including microwaves and optics. Although optical antennas are usually fed by either a single source or many sources incoherently, recent studies demonstrate an ability of resonant nanostructures to cause the synchronization of quantum source spontaneous emission, i.e., Dicke superradiance effect. What remains poorly explored is how the coherent excitation can affect antenna performance: its multipole composition, directivity, efficiency, and Purcell effect. In this paper, we investigate an antenna excited by two coherent sources and demonstrate that this approach can boost antenna performance. To make the discussion independent of the frequency range, we restrict the consideration by all-dielectric antennas, which attract significant interest in both microwaves and optics. We explore that coherent excitation of a dielectric antenna by two dipole sources makes it feasible to control the multipoles resulting in boosted superradiance and enhanced directivity. Interestingly, it makes possible excitation of nonradiative field configuration, anapole state, and turning it on/off on demand. Remarkably, this approach also allows reducing the quenching effect and enhancing the radiation efficiency without changing the antenna's geometry. We design a dielectric subwavelength coherently driven antenna operating in both superdirective and superradiative regimes simultaneously with the total enhancement factor over $2 \cdot 10^3$.

# Introduction

Antennas are a crucial element for many vital wireless technologies, including communications and power transfer[1]. Being dictated by applications, a plethora of antennas in the radio and microwave frequency ranges have been invented, including microstrip antennas[2], reflector antennas[1,3], dielectric antennas[4,5] to mention just a few. More recently, the optical counterpart, a nanoantenna, has also been invented for quantum optics, spectroscopy, and communications on a chip[6–12]. First attempts to nanoantennas developing were based on utilizing the plasmonic response of noble metals which quickly faced with the inevitable dissipative losses interfering in many thermo- and loss-sensitive applications. All-dielectric nanoantennas have been suggested to address the material losses issue and are becoming indispensable for sensing and spectroscopy [9,13–17] and as constituent elements for metasurfaces and metamaterials[18–21].

Typically, nanoantennas are fed by a single quantum optical source (molecule, QD) or by many sources *incoherently*. In this scenario, the antenna-effect consists in the enhancement of the source emission via Purcell effect, i.e., increasing of the radiative decay rate of a source induced by the enlarged local density of optical states (LDOS)[22–30]. Despite that the Purcell effect can lead to significant enhancement of the emitted power ($P_{rad}$)[31–33], it scales with the number of quantum sources ($N$) as $P_{rad} \propto N$, due to the incoherent nature of spontaneous emission.

In 1954, Robert Dicke theoretically demonstrated[34] that in a system of $N$ excited two-level atoms, the spontaneous emission can become correlated. In result, the entire system radiates as a source with a dipole $d \sim Nd_0$ ($d_0$ is the dipole moment of a single atom), and hence scales as $P_{rad} \propto N^2$. In the time-resolved scenario, it leads to an increase in emission rate and narrowing of the emitted pulse [35,36]. The synchronization of spontaneous emission can arise in atom ensembles confined in a subwavelength region of volume $< (\lambda)^3$, where $\lambda$ is the radiation wavelength. Interestingly, resonant optical nanostructures can assist in achieving of correlated spontaneous emission of coupled sources, located not necessarily at small mutual distances [37–43]. The Dicke radiation has been predicted and observed for $N = 2$ sources[44] and ensembles of many $N \gg 1$ sources[36] in a variety of systems, including atoms[45,46], ions[44], quantum dots[36], qubits[47], and Josephson junctions[48].

The superradiance effect has an analogy in classical electrodynamics and antenna theory and may consist in the matching of coherently excited closely arranged antennas (i.e., antenna arrays)[49] with an excitation channel (e.g., waveguide) or enlarging of elastic scattering in arrays of (nano) particles[50] and on-chip photonic crystals[51]. Also, the effect plays an essential role in Josephson-junction arrays[48] and utilized for the emission of highly intense Cherenkov pulses[52]. In analogy with its quantum counterpart, rapid superlinear enhancement of the emission power with the number of antennas or scatterers occurs in this regime. A good discussion on the classical analogy of the Dicke radiation can be found in Refs.[40,53].

Hence, the radiation of an array of coherent sources plays an essential role in modern science and technologies across the entire spectrum. However, to the best of our knowledge, the question of how the coherent excitation can affect antenna performance (multipole composition, directivity, efficiency, and Purcell effect) remains unexplored. In fact, the simultaneous excitation of an antenna by several coherent sources is expected to alter its electromagnetic properties drastically via tailoring of its multipole composition. In this paper, we take the first attempt to study this issue and demonstrate that this approach can boost antenna performance. Firstly, we show that coherent excitation of an antenna by two localized sources [Figure 1(a)] makes it feasible to *control the excitation of multipoles* and as a result, its electromagnetic properties. It leads to the ability of coherent tuning of radiated power from almost zero values (subradiance) to significantly enhanced (superradiance). Interestingly, it makes possible excitation of anapole state and turning it on/off at will. Further, we explore that this approach allows reducing the quenching effect and *strengthening the radiation efficiency* at some specific phase via coherent avoidance of higher mode excitation without changing the geometry of the antenna. Finally, we demonstrate that utilizing this approach allows designing an antenna operating in *superdirective and superradiance regimes simultaneously*, Figure 1(b). To make the discussion independent of the frequency range, we restrict the consideration by all-dielectric antennas, which attract significant interest for both microwaves[32,54–56] and optics[8,17,57,58] and, hence, use dimensionless units.

We also note that the coherent effects in optics attract a great interest today because they have been demonstrated to be allowed to achieve perfect absorption[59–62], ultimate all-optical light manipulation[63], and enhance wireless power transfer[64,65]. In this work, we show that the coherent excitation brings new exciting and useful effects to antenna science that otherwise are unattainable or difficult to achieve in traditional approaches.

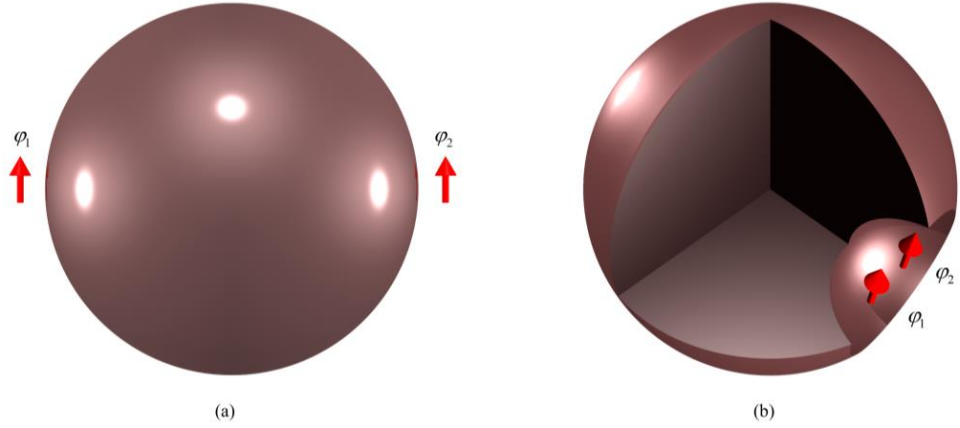

**Figure 1**. Schematic representation of considered dielectric coherently driven (a) simple antenna and (b) notched superdirective antenna. The direction of the dipoles is along the y-axis.

## Results and discussion

**Sub- and superradiance.** – We begin our analysis with consideration of a simple symmetric system depicted in Figure 1(a). The structure consists of a high-index dielectric resonator with radius $R$ and refractive index of $n=4$. The value of the refractive index is selected based on the typical values of the materials used in optics and microwaves[54,66,67]. Such a dielectric resonator supports Mie modes of a different order [see Supplementary Materials (SM)#I for details]. The first mode, magnetic dipole (md), is formed when the radius of the resonator satisfies $R \approx \lambda/2n$, or for fixed $n=4$, at $\lambda/R \approx 8$ [18,21]. Next resonant modes, electric dipole (ed), magnetic quadrupole (mq) etcetera, appear at shorter wavelengths.

It is illustrative to consider a simple but instructive scenario of the resonator excitation by two oppositely directed plane waves. We restrict ourselves by the spectral range where the resonator supports only electric and magnetic dipole resonant modes. Upon excitation by two plane waves, due to linearity of the problem, the absolute values of Mie scattering dipole electric and magnetic amplitudes can be expressed as $|a_1 + a_1 e^{i\varphi}|$ and $|b_1 - b_1 e^{i\varphi}|$, respectively (see SM#I for details). The different signs stem from the pseudo-vector character of the magnetic dipole. Thus, we conclude that even such a simple scenario opens an opportunity to coherently tune scattering from the resonator with turning on/off its modes at specific relative phases ($\varphi$) of the waves (see SM#I for details).

Now we assume the resonator is excited by two dipoles of equal amplitude ($P_{dy}$, $|P_{dy}|=1$) but different phases ($\varphi_1$ and $\varphi_2$), and polarized along the y-axis, as shown in Figure 1(a). In the dipole approximation, the system can be considered analytically by the discrete dipole approximation (DDA) approach[57,68]. Following this approach, the resonator is described as a superposition of the md and ed moments with the magnetic ($\alpha_p^m$) and electric ($\alpha_p^e$) polarizabilities and corresponding magnetic ($M_p$) and ($P_p$) dipole moments applied to the resonator's center. The solution of the DDA equations (see SM#II for details) yields the following nonzero components of the electric and magnetic dipole moments

$$P_{py} = \alpha_p^e A_{pd} P_{dy} (e^{j\varphi_1} + e^{j\varphi_2}), \tag{1}$$

$$M_{pz} = \alpha_p^m \sqrt{\frac{\varepsilon_0}{\mu_0}} D_{pd} P_{dy} (e^{j\varphi_1} - e^{j\varphi_2}), \tag{2}$$

where $\varepsilon_0$ and $\mu_0$ are the permittivity and permeability of free space respectively; $A_{pd}$ and $D_{pd}$ are coefficients (see SM#II). The analysis of these equations shows that when the dipole sources radiate in-phase ($\varphi_1 - \varphi_2 = 0$), the magnetic dipole moment of the resonator vanishes, $M_{pz} = 0$, while the electric dipole moment is maximal ($|P_{py}|=2|\alpha_p^e A_{pd}|$) and twice larger than in the case of excitation by a single source. In the case of opposite phase excitation ($|\varphi_1 - \varphi_2|=180$ deg), the situation is reversed, and the resonator possesses zero electric dipole moment and enhanced magnetic dipole ($|M_{pz}|=2\sqrt{\varepsilon_0/\mu_0}|\alpha_p^m D_{pd}|$). Thus, the two-dipole coherent excitation makes it feasible to tailor the multipole composition of a resonator.

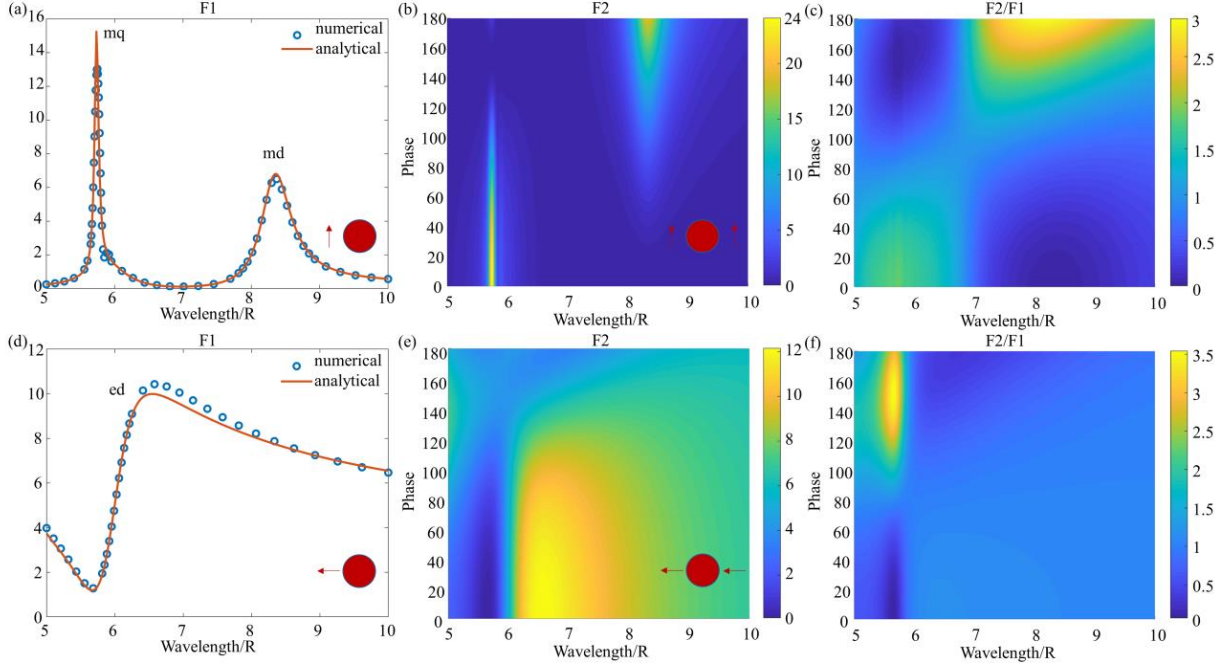

**Figure 2.** Purcell factor of a single dipole source ($F_1$) with (a) tangential (TD) and (d) longitudinal (LD) orientation vs. wavelength normalized to the radius ($\lambda/R$) calculated both numerically and analytically. (b), (e) Collective radiation enhancement of two dipoles ($F_2$) with (b) TD and (e) LD orientation vs. $\lambda/R$ and phase difference of dipoles ($\varphi_d = |\varphi_1 - \varphi_2|$). (c), (f) Normalized collective radiation enhancement ($F_2/F_1$) for (c) TD and (f) LD orientation vs. $\lambda/R$ and $\varphi_d$.

To explore how this coherent antenna excitation affects the antenna properties, we consider the Purcell effect in this system. Firstly, we calculate the Purcell factor ($F_1$) for the single dipole source for both tangential (TD) and longitudinal (LD) orientation, Figures 2(a), (d). For numerical calculation, we use the input-impedance approach reported in Ref.[29]. Note that since this system is assumed to be lossless, this approach coincides with the ratio of power radiated in the far zone to the same in the absence of resonator[29]: $F_1 = P_{rad}/P_{0,rad}$, where $P_{rad}$ is the radiated power for the presence of resonator and $P_{0,rad}$ is for free space. The results of numerical calculation of the Purcell factor of a single dipole ($F_1$) for tangential (TD) and longitudinal (LD) orientation in CST Microwave Studio are presented in Figure 2(a) and (d) by circles. We use analytical calculations based on Green's function approach for the verification of our numerical results[69,70]. The numerical and analytical results are in an excellent agreement. For exact geometrical parameters used for calculations see *Methods*.

For TD polarization, we observe two resonant modes, which are magnetic dipole (md) and magnetic quadrupole (mq) at $\lambda/R=8.37$ and $\lambda/R=5.75$ respectively [Figure 2(a)]. The electric dipole (ed) moment is not excited in TD polarization due to its zero overlap with the source, while effectively excited at $\lambda/R=6.58$ for LD polarization [Figure 2(d)]. These excited resonant modes lead to a dramatic increase in power radiation and Purcell factor for both TD and LD polarisations. This result coincides with our analytical result (red curve) and with results reported in earlier works[18,71].

Now we show that although the high Purcell effect is achieved for one dipole, it can be further increased by introducing another dipole near the resonator, Figure 1(a). We define the *collective radiation enhancement* $F_2$ as

$$F_2(\varphi_d) = \frac{P_{\text{rad}}(\varphi_d)}{P_{0,\text{rad}}(\varphi_d)}, \quad (3)$$

where $P_{\text{rad}}(\varphi_d)$ [$P_{0,\text{rad}}(\varphi_d)$] is the total radiated power of the two dipoles excited with the phase difference ($\varphi_d$) with [without] the resonator. The numerical results of collective radiation enhancement of two dipoles ($F_2$) versus $\lambda/R$ and phase difference ($\varphi_d$) are summarized in Figures 2(b) and (e). We see that $F_2$ is raised to 24 at mq ($\lambda/R=5.75$) and 20 at md ($\lambda/R=8.37$) resonances at certain phases, which is ~85% and ~300% increase compared to $F_1$ for TD [Figure 2(b)]. On the other hand, $F_2$ reaches 12 for ed of LD orientation, which is ~20% enhancement [Figure 2(e)]. The results for md and ed coincide with the above analytical DDA description.

The ratio $F_2/F_1$ gives the normalized collective radiation enhancement. If $F_2/F_1>1$ ($F_2/F_1<1$), the antenna boosts (suppresses) the coherent collective emission from the sources. The results presented in Figures 2(c), (f) show that the antenna boosts collective emission at certain phases. Remarkably, at $\lambda/R=5.68$ where $F_1$ is minimum, an additional dipole with $\varphi_d=150$ deg can increase the power radiation 3.5 times higher [Figure 2(f)], which is associated with an anapole state and discussed below.

Besides collective radiation enhancement, manipulating the phase difference can also achieve subradiance effect, which technically "turns off" the antenna without changing the structure. As previously mentioned, $F_2$ can be maximized up to 24 at $\lambda/R=5.75$ for TD. In contrast, $F_2$ can also be lower down to ~0 by introducing a relative phase of $\varphi_d=155$ deg. Besides, while $F_2$ features a 300% boost at $\lambda/R=8.37$ for $\varphi_d=180$ deg, zero power radiation can be achieved by altering the phase of dipoles to be the same. A similar effect is also noticed at $\lambda/R = 5.68$ for LD. These results indicate that one could easily tune the Purcell factor by just adjusting the phase difference without changing the structure of the antenna.

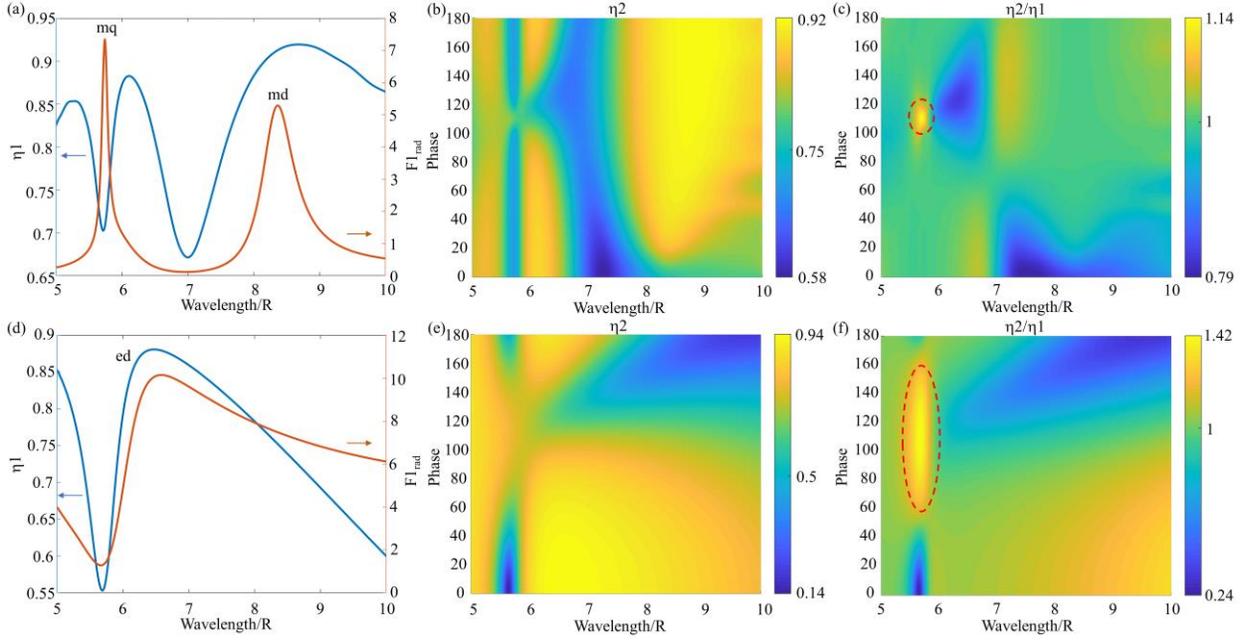

**Figure 3.** Radiation efficiency and radiative Purcell factor of the antenna consisting of dipole and resonator with radius R and refractive index, $n = \sqrt{16 + 0.1i}$, depending on $\lambda/R$ and $\varphi_d$. (a) and (d) show both radiated efficiency (blue curve) and radiative Purcell factor (red curve) of one dipole for TD and LD orientation respectively depending on $\lambda/R$. (b), (e) Radiation efficiency of two dipoles ($\eta_2$) with TD and LD orientation respectively depending on $\lambda/R$ and $\varphi_d$. Normalized radiation efficiency ($\eta_2/\eta_1$) for (c) TD and (f) LD orientation depending on $\lambda/R$ and $\varphi_d$. The value of $\eta_2/\eta_1$ is maximum at ~5.7 with 110 deg phase difference for both TD and LD orientation. The red dashed circle shows the region where the maximum $\eta_2/\eta_1$ is achieved.

**Efficiency boosting.** – Besides enhancement of radiated power and radiative decay rate, another essential quantity characterizing any antenna is its radiation efficiency ($\eta_1$), defined as $\eta_1 = P_{rad}/P_{tot}$, where $P_{tot}$ is the total delivered power to the system. Note that although this definition is fair for both microwave and optics, in quantum optics another related definition (quantum efficiency) through the number of radiated photons ($N_{rad}$) is common: $\eta_1 = N_{rad}/N_{tot}$, where $N_{tot}$ is the total number of excited quasiparticles (electrons, excitons).

The radiation efficiency of a source, e.g., molecule or quantum dot (QD) can be increased via the Purcell effect[72] by speeding up the radiative decay rate and reducing the nonradiative decays. This scenario can boost radiation efficiency of a dipole source arranged nearby a resonator with a resonant mode. However, in reality, the emission of a dipole source located closely to a resonator surface gets dissipated because of the quenching effect, i.e., strong dissipation through excitation of nonresonant higher-order modes[73–75].

Here, we show that the *coherent excitation by two dipole sources can lead to boosting of radiation efficiency* owing to weakening of other modes besides the one of our interest. To this end, we introduce a realistic imaginary part to the resonator refractive index, $n = \sqrt{16 + 0.1i}$. Firstly, we calculate the radiated efficiency of one dipole ($\eta_1$) for TD and LD orientations, shown in Figures 3(a), (d) by red curves. The radiation efficiency at the md resonance of TD orientation is ~0.7 and at the ed resonance of LD orientation is ~0.85. In consequence, the antenna dissipates a significant amount of power before radiating, although it may have a high Purcell factor, Figures 3(a), (d), blue curves.

Further, we define the radiation efficiency of the two-dipole excitation scenario as

$$\eta_2(\varphi_d) = \frac{P_{\text{rad}}(\varphi_d)}{P_{\text{tot}}(\varphi_d)}, \qquad (4)$$

where $P_{\text{tot}}(\varphi_d)$ is the total delivered power to the system, which now is the relative phase-dependent value. The results of the numerical calculation of this coherent radiative efficiency as a function of the dimensionless wavelength and the relative phase are presented in Figures 3(b), (e). To compare these results with $\eta_1$, we take their ratio, presented in Figures 3(c), (f). We see that the presence of the second dipole source can increase the overall radiation efficiency ($\eta_2 / \eta_1 > 1$) for both TD and LD orientations at certain relative phases. We observe the most significant effect at $\lambda / R = 5.7$ for both TD and LD. The efficiency of TD at mq is increased by 14% at the relative phase of 110 deg, as shown in Figure 3(c), which reduces the dissipation and hence elevates the power radiation. On the other hand, at $\lambda / R = 5.7$ of LD, the antenna that originally had a low efficiency of 55% [Figure 3(d)] has become much more efficient with gain 42% [Figure 3(f)] in the collective excitation scenario. In short, these results convincingly show that the collective coherent excitation does not only have the advantage of increasing the Purcell effect but also the efficiency, which is the most crucial characteristic for any antenna.

Let us consider the LD polarized excitation at $\lambda/R = 5.7$ more detailed. Figures 4(a) and (b) demonstrate the dependences of collective radiation enhancement ($F_2$) and mutual radiation efficiency ($\eta_2$) on wavelength for the relative phase of 0 deg (blue curves) and 110 deg (red curves). We observe that both characteristics get increased at 110 deg of phase, and hence the antenna at this wavelength not only radiates more but also does it in a much more efficient way. To understand why this happens, we show the electric field distribution at $\lambda/R = 5.7$ for both phases, Figure 4(c). The vector E-field distribution at the phase of 0 deg reminds that of the anapole state[76] with enhanced linear distribution corresponding to the Cartesian dipole and two loops, corresponding to the toroidal moment[76,77]. Also, this point corresponds to the well-known anapole state of a spherical particle manifesting itself in the vanishing of Mie scattering amplitude $a_1$ and drop in the scattering efficiency (see SM#I for details). The anapole character of the excited state leads to reducing of $F_2$. Importantly, this anapole state is associated with increased E-field in the center, which leads to increased dissipation losses in the center, Figure 4(d).

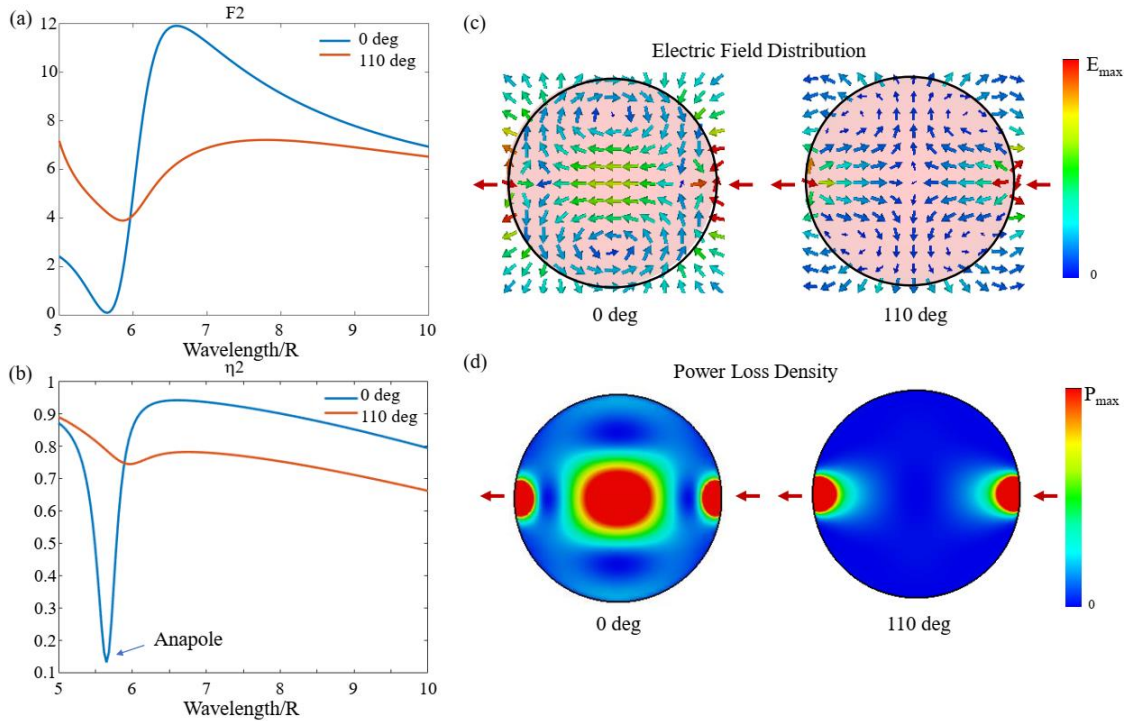

**Figure 4.** (a), (b) Dependences of the collective radiation enhancement ($F_2$) and mutual radiation efficiency ($\eta_2$) on wavelength for the relative phase of 0 deg (blue curves) and 110 deg (red curves)

for the longitudinal (LD) excitation. (c), (d) E-field and power loss density in the resonator for the zero phase difference (left column) and 110 deg (right column) at $\lambda/R = 5.7$.

Figures 4(c) and (d), right column, show that at the relative phase of 110 deg, the coherent excitation leads to *"turning off" the anapole state* with suppression of the E-field in the resonator center with the corresponding suppression of the power loss density and gain in both collective radiation enhancement ($F_2$) and mutual radiation efficiency ($\eta_2$) at $\lambda/R = 5.7$.

**Superdirectivity.** – The above analysis shows that the coherent excitation of an antenna by (at least) two sources leads to enhancement of the radiated power and boosting of the radiation efficiency. The reason lying behind these effects is the tailoring of excited multipoles in the antenna. In this section, we demonstrate that the same approach provides a powerful tool for the antenna directivity tailoring. Namely, we show that at a certain relative phase, an antenna can operate in both superradiation and superdirectivity regimes simultaneously. To this end, we utilize the recently reported design for all-dielectric superdirective notched antenna[16,56], Figure 5(a). Superdirective antennas are those whose size is smaller than the operation wavelength in all three directions, and directivity is much larger than the directivity of a short dipole antenna ($D_{max} = 1.5$)[16,56,78–83]. The superdirective antenna operation relies on creating rapidly spatially oscillating currents in a subwavelength area, which leads to excitation of higher multipoles[56,78]. If the multipoles are excited with right certain phases and amplitudes, their far-fields interfere, forming a spatially narrow radiation beam. In result, the antenna becomes very directive despite its subwavelength volume. Here we demonstrate that the excitation of a superdirective antenna by two coherent sources allows ultimate control of excited multipoles and radiation pattern.

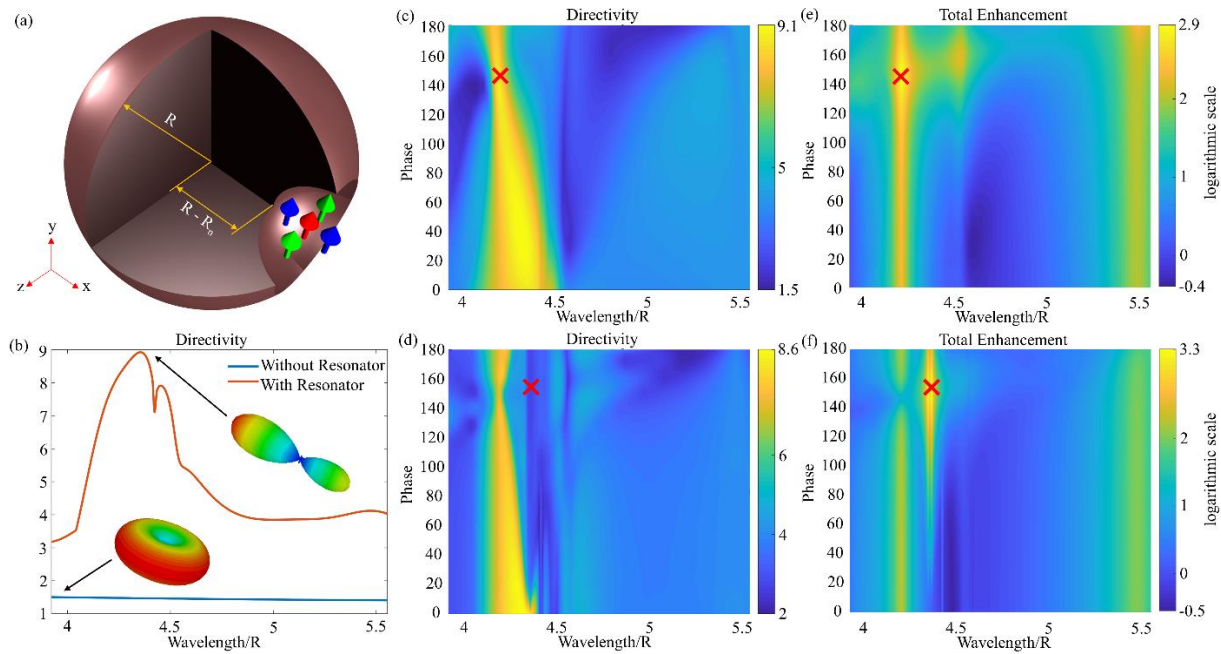

**Figure 5.** (a) Schematic representation of the notched antenna comprising two coherent dipoles of different configuration. (b) Directivity of the single dipole [red arrow in (a)] with and without resonator depending on $\lambda/R$. Insets show the corresponding radiation patterns. (c), (d) Directivity of two dipoles placing along the (c) x-axis [blue arrows in (a)] and (d) z-axis [green arrows in (a)] in the notch. (e), (f) Total enhancement, $\Sigma = D_{max} \cdot F_2$, in logarithmic scale for two dipoles placing along the (e) x-axis and (f) z-axis in the notch of the resonator. The red crosses show the points of maximum total enhancement that results in superradiative and superdirective regimes simultaneously.

Figure 5(b) presents the results of directivity calculation of the notched antenna for single dipole excitation [Figure 5(a), red arrow]. Following the antenna textbooks, we define the directivity as $D_{max} = 4\pi P_{max}(\theta,\varphi)/P_{rad}$, where $(\theta,\varphi)$ are the angular coordinates of the spherical coordinate system, and $P_{max}$ is the power in the direction of the main lobe. This value is normalized so that the isotropic point source has $D_{max}=1$ and the dipole source has $D_{max}=1.5$. In our system, we observe the maximum directivity $D_{max}=9$ at $\lambda/R=4.35$ (red curve), which is much higher than that in free space (~1.5, blue curve) and satisfies the definition of the superdirectivity regime. The insets demonstrate that the three-dimensional radiation pattern can be concentrated in a single direction in the presence of the designed antenna. Thus, these results show that the superradiative antenna significantly increases the angular density of the radiated power. This idea has been theoretically suggested for optics in Ref.[16] and experimentally realized in microwaves in Ref.[56].

The presence of the second dipole source allows coherent tuning of this superdirective antenna. For the arrangement of the sources along the x-axis [Figure 5(a), blue arrows], we observe the preservation of the superdirectivity regime with the maximum directivity (~9.1), which changes with the relative phase, Figure 5(c). To achieve a superradiative and superdirective antenna, we compute the total enhancement, $\Sigma = D_{max} \cdot F_2$, that is the product of directivity and collective radiation enhancement, $F_2$ (see SM#V for details). For this x-axis orientation, the maximum overall improvement of 880 in linear scale (2.9 in logarithmic scale) is achieved, Figure 5(e, red cross). At this point, the antenna possesses both superdirectivity and superradiation effects with $F_2 = 100$ and $D_{max} = 8.5$. When the sources are placed along the z-axis [Figure 5(a), green arrows], a higher maximum total enhancement of 2160 in linear scale (3.3 in logarithmic scale) is observed for z-axis orientation despite smaller directivity (~2.7) in this case, and the total enhancement is mainly contributed by $F_2 = 785$ in linear scale (see SM#V for details). Thus, for the x-axis arranged coherent sources, the antenna can operate in both superradiative ($F_2 / F_1 > 1$) and superdirective ($D_{max} \gg 1$, $\lambda / R > 1$) regimes simultaneously.

## Conclusions

In this paper, we have explored the issue of how the coherent excitation by several sources can affect antenna performance. We have shown that coherent excitation of an antenna by two localized sources makes it feasible to control the excitation of multipoles and as a result, its electromagnetic properties. It leads to the ability of coherent tuning of radiated power from almost zero values (subradiance) to significantly enhanced (superradiance). We have explored that this approach allows reducing the quenching effect and strengthening the radiation efficiency at some specific phase via coherent avoidance of higher mode excitation without changing the geometry of the antenna. The approach makes feasible excitation of nonradiative field configuration, anapole state, in the spherical antenna and turning it on/off on our demand. We have also demonstrated that utilizing this approach allows designing an antenna operating in superdirective and superradiance regimes simultaneously with the total enhancement factor over $2 \cdot 10^3$. We believe that the general effects reported in this study will found an application in a broad range of technologies, including coherently driven antennas and active nanophotonics.

## Methods

*Numerical simulations* are performed in commercial software CST Microwave Studio 2018. CST Microwave Studio is a full-wave 3D electromagnetic field solver based on finite-integral time-domain solution technique. A nonuniform mesh was used to improve accuracy in the vicinity of the dielectric resonators where the field concentration was significantly large. The dielectric permittivity is indicated in the text. The exact geometrical parameters used for simulations are as follows. The excitation dipole sources have the same length $l_d$ = 20 nm. Fig. 1(a): the resonator radius is R = 60 nm; the dipoles are located at the distance of 10 nm from the resonator surface as shown. Fig. 1(b): the notched antenna has the radius R = 90 nm and a notch with radius $R_n$ = 40 nm. In Ref.[16] it has been shown that this geometry provides the regime of largest directivity. The center of the notch is exactly on the surface of resonator, and the midpoint between the dipoles is 20 nm away from the surface. Since the parameters are scalable, the results are discussed in the dimensionless units (wavelength normalized to the radius of the resonator) throughout the paper as they can be used in both microwave and optical wave application.